\documentclass[cits]{PoS}

\usepackage{textcomp}
\usepackage[latin1]{inputenc}
\usepackage{amsfonts}
\usepackage{amssymb}
\usepackage{amsbsy}
\usepackage{amsmath}
\usepackage{xspace}
\usepackage{graphicx} 

\usepackage{caption}
\newcommand{\FourS}{\Upsilon(4S)}
\newcommand{\FiveS}{\Upsilon(5S)}

\newcommand{\invfb}{\,{\rm fb}^{-1}}

\newcommand{\mbc}{{M_{\textrm{bc}}}}
\newcommand{\deltae}{{\Delta E}}
\newcommand{\BR}{{\mathcal B}}

\newcommand{\lint}{{L_{\textrm{int}}}}

\newcommand{\gev}{{\hbox{GeV}}}

\newcommand{\bs}{{B_s^0}}
\newcommand{\bsst}{{B_s^{\ast}}}
\newcommand{\barbsst}{{{\bar B_s}^{\ast}}}
\newcommand{\bsST}{{B_s^{(\ast)}}}
\newcommand{\bsSTbsST}{{\bsST{\bar B_s}^{(\ast)}}}
\newcommand{\KS}{{K_S^0}}
\newcommand{\KL}{{K_L^0}}
\newcommand{\ds}{{D_s^-}}

\newcommand{\bsdspi}{{\bs\to D_s^-\pi^+}}
\newcommand{\bsdsstpi}{{\bs\to D_s^{\ast-}\pi^+}}

\newcommand{\fss}{{\left(90.1^{+3.8}_{-4.0}\pm0.2\right)\%}}

\newcommand{\bfbstodsstpi}{(
2.4{}^{+0.5}_{-0.4}({\rm stat.})\pm0.3({\rm syst.})\pm0.4(f_s))\times10^{-3}}
\newcommand{\bfbstodsrho}{(
8.5{}^{+1.3}_{-1.2}({\rm stat.})\pm1.1({\rm syst.})\pm1.3(f_s))\times10^{-3}}
\newcommand{\bfbstodsstrho}{(13.0{}^{+2.3}_{-2.1}({\rm stat.})
\pm1.7({\rm syst.})\pm1.7({\rm pol.})\pm1.9(f_s))\times10^{-3}}

\newcommand{\bsdsrho}{{\bs\to D_s^-\rho^+}}
\newcommand{\bsdsstrho}{{\bs\to D_s^{\ast-}\rho^+}}

\newcommand{\jpsi}{{J\!\!/\!\!\psi}}

\newcommand{\bsjpsieta}{{\bs\to\jpsi\,\eta}}
\newcommand{\bspipi}{{\bs\to\pi^+\pi^-}}
\newcommand{\bskpi}{{\bs\to K^+\pi^-}}
\newcommand{\bskk}{{\bs\to K^+K^-}}
\newcommand{\bskzkz}{{\bs\to K^0\bar K^0}}

\newcommand{\bfbstojpsieta}
{{(3.69\pm0.95({\rm stat.}){}^{+0.65}_{-0.95}({\rm syst.}))\times10^{-4}}}
\newcommand{\bfbstokk}{{(3.8{}^{+1.0}_{-0.9}({\rm stat.})\pm0.7({\rm syst.}))\times 10^{-5}}}
\newcommand{\bfbstokpi}{{2.6\times10^{-5}}}
\newcommand{\bfbstopipi}{{1.2\times10^{-5}}}
\newcommand{\bfbstokzkz}{{6.6\times 10^{-5}}}

\title{$\mathbf{\bs}$ Decays at Belle}

\ShortTitle{$\bs$ Decays at Belle}

\author{\speaker{Remi Louvot}\\
  (On behalf of the Belle collaboration)\\
  Laboratoire de Physique des Hautes \'Energies,\\
  \'Ecole Polytechnique F\'ed\'erale de Lausanne~(EPFL), Lausanne, Switzerland\\
  E-mail: \email{remi.louvot@epfl.ch}}

\abstract{
  The large data sample being recorded with the Belle detector at the
  $\FiveS$ energy provides a unique opportunity to study the poorly-known
  $\bs$ meson.
  Three analyses, using a sample of 23.6~$\invfb$, are presented.
  We report the first observation of the three domi\-nant exclusive $\bs$ decays,
  $\bsdsstpi$, $\bsdsrho$ and $\bsdsstrho$,
  the first observation of the $CP$-eigenstate decay $\bsjpsieta$, and
  the results from a study of the charmless $\bskk$, $\bskpi$, $\bspipi$ and $\bs\to\KS\KS$ decays. 
  The following branching fraction measurements are presented:
  (When mentioned, the systematic errors due to the uncertainties on the fraction $f_s=\sigma(e^+e^-\to\bsSTbsST)/\sigma_{b\bar b}$ and on the decay polarization (pol.) are quoted separately.)
  $$\BR(\bsdsstpi)=\bfbstodsstpi\,,$$
  $$\BR(\bsdsrho)=\bfbstodsrho\,,$$
  $$\BR(\bsdsstrho)=\bfbstodsstrho\,,$$
  $$\BR(\bsjpsieta)=\bfbstojpsieta~\rm{and}$$
  $$\BR(\bs\to K^+K^-)=\bfbstokk\,,$$ 
  and the limits at $90\%$ confidence level:
  $$\BR(\bs\to\pi^+\pi^-)<\bfbstopipi\,,$$
  $$\BR(\bskpi)<\bfbstokpi~\rm{and}$$
  $$\BR(\bskzkz)<\bfbstokzkz\,.$$

 ~

\today

}

\FullConference{
  The 2009 Europhysics Conference on High Energy Physics,\\
  July 16 - 22, 2009\\
  Krakow, Poland}

\begin{document}

\section*{Introduction}

The Belle experiment \cite{NIMA_479_117}, located at the interaction point of
the KEKB asymmetric-energy $e^+e^-$ collider \cite{NIMA_499_1},
was designed for the study of $B$ mesons\footnote{The notation ``$B$'' refers either to a $B^0$ or a $B^+$.
  Moreover, charge-conjugated states are implied everywhere.}
produced in $e^+e^-$ annihilation at a center-of-mass (CM) energy corresponding to the mass of
the $\FourS$ resonance ($\sqrt s\approx10.58~\gev$).
After having recorded an unprecedented sample of $\sim800$ millions of $B\bar B$
pairs,
the Belle collaboration started to record collisions at higher energies,
opening the possibility to study other particles, like the $\bs$
meson.
Up to now, $\lint=23.6~\invfb$ of data have been analyzed at the energy of the $\FiveS$ resonance
($\sqrt s\approx10.87~\gev$).

The $\FiveS$ resonance is above the $\bs\bar\bs$ threshold and it was naturally
expected that the $\bs$ meson could be studied with $\FiveS$ data as well as
the $B$ mesons are with $\FourS$ data.
The large potential of such $\FiveS$ data was quickly confirmed
\cite{PRL_98_052001,PRD_76_012002} with the 2005 engineering run
representing 1.86~$\invfb$.
The main advantage with respect to the hadronic colliders is
the possibility of measurements of absolute branching fractions.
However, the abundance of $\bs$ mesons in $\FiveS$ hadronic events has to be
precisely determined.
Above the $e^+e^-\to u\bar u, d\bar d, s\bar s, c\bar c$ continuum events,
the $e^+e^-\to b\bar b$ process can produce different kinds of final states:
seven with a pair of non-strange $B$ mesons ($B^{\ast}\bar B^{\ast}$,
$B^{\ast}\bar B$, $B\bar B$, $B^{\ast}\bar B^{\ast}\pi$, $B^{\ast}\bar B\pi$,
$B\bar B\pi$ and $B\bar B\pi\pi$),
three with a pair of $\bs$ mesons
($\bsst\barbsst$, $\bsst\bar\bs$ and $\bs\bar\bs$),
and final states involving a bottomonium resonance below the $B\bar B$
threshold \cite{PRL_100_112001}.
The $B^{\ast}$ and $\bsst$ mesons always decay by emission of a photon.
The total $e^+e^-\to b\bar b$ cross section at the $\FiveS$ energy was measured
to be $\sigma_{b\bar b}=(302\pm14)$~pb
\cite{PRL_98_052001,PRD_75_012002} and the fraction of $\bs$ events to be
$f_s=\sigma(e^+e^-\to\bsSTbsST)/\sigma_{b\bar b}=(19.3\pm2.9)$~\%
\cite{PLB_667_1}.
The dominant $\bs$ production mode, $b\bar b\to\bsst\barbsst$,
represents $f_{\bsst\barbsst}=\fss$ of the $b\bar b\to\bsSTbsST$ events
\cite{PRL_102_021801}.

For all the exclusive modes presented here, the $\bs$ candidates are fully reconstructed
from the final-state particles.
From the reconstructed four-momentum in the CM, $(E_{\bs}^{\ast},p_{\bs}^{\ast})$,
two variables are formed:
the energy difference $\deltae=E_{\bs}^{\ast}-\sqrt s/2$ and the
beam-constrained mass $\mbc=\sqrt{s/4-p_{\bs}^{\ast2}}$.
The signal coming from the dominant $e^+e^-\to\bsst\barbsst$ production mode
is extracted from a two-dimensional fit performed on the distribution of these
two variables.
The corresponding branching fraction is then extracted using the total efficiency (including sub-decay branching fractions)
determined with Monte-Carlo simulations, $\sum\varepsilon\BR$,
and the number of $\bs$ mesons produced via the $e^+e^-\to\bsst\barbsst$ process, $N_{\bs}=2\times\lint\times\sigma_{b\bar b}\times f_s\times f_{\bsst\bar\bsst}=(2.48\pm0.41)\times10^6$.


\section{Observation of three new CKM-favored $\mathbf{\bs}$ decays}

\begin{figure}[!h]
\begin{minipage}{0.65\linewidth}
\includegraphics[width=\linewidth]{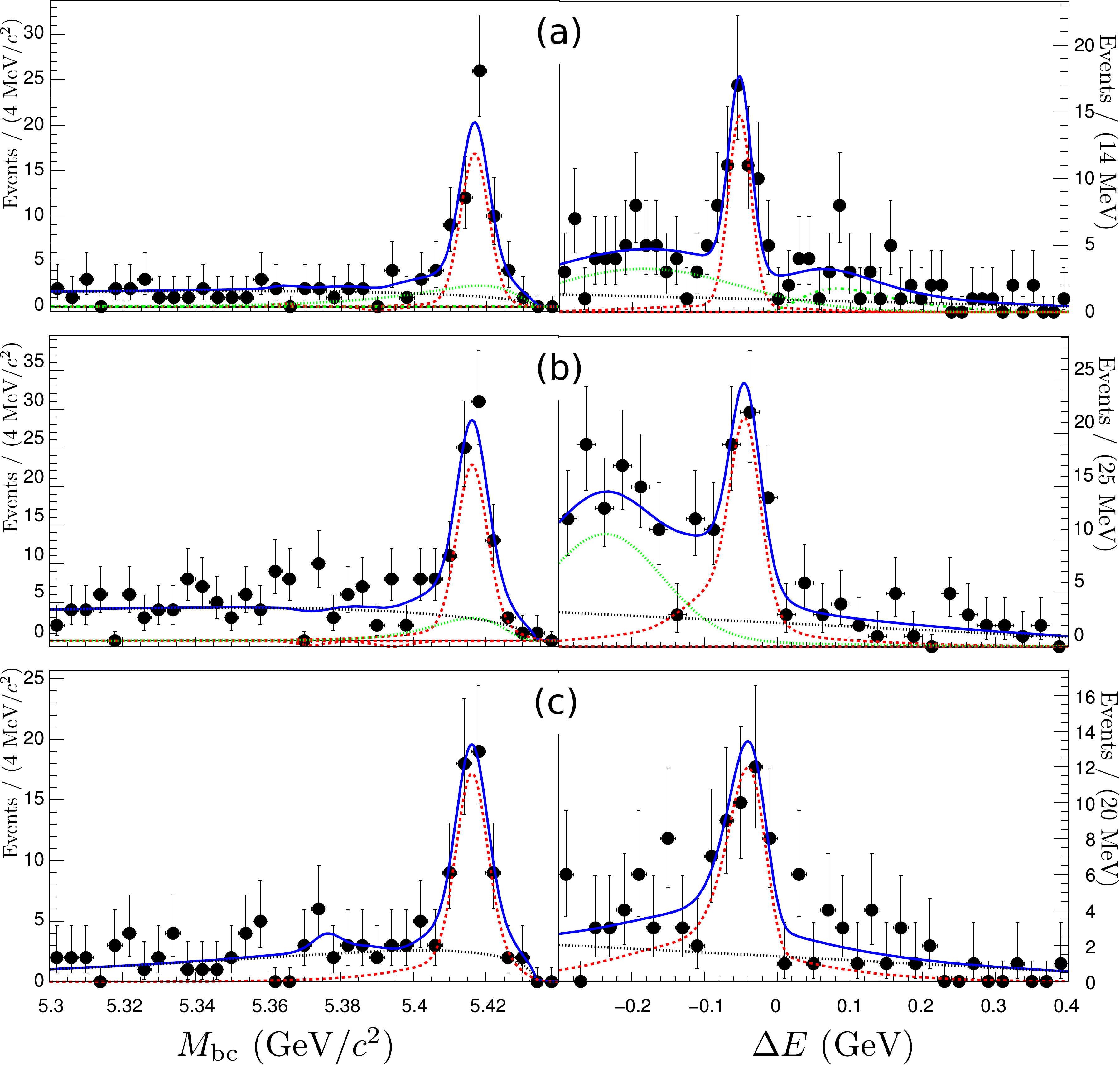}
\end{minipage}~~
\begin{minipage}{0.32\linewidth}
\caption[]{\label{fig:dominant}
  Histograms with the fit results.
  The left (right) plots shows the $\mbc$ ($\deltae$) projections for the candidates with $\deltae$ ($\mbc$)
  in the $\bsst\bar\bsst$ signal region,
  The red-dashed (green-dotted, black-dotted, solid-blue) lines represent the signal
  (peaking $\bs$ background, continuum, total) fitted PDF.
  (a): $\bsdsstpi$ candidates. $\bsdspi$ ($\bsdsrho$) events can be distinguished on the right (left) of the signal peak in the $\deltae$ plot.
  (b): $\bs\to D_s^-\rho^+$ candidates. The peaking background comes from $\bs\to D_s^{\ast-}\rho^+$ events. (c): $\bs\to D_s^{\ast-}\rho^+$ candidates.}
\end{minipage}
\end{figure}

Following our recent measurement of $\bsdspi$ \cite{PRL_102_021801},
we present for the first time an extension of this analysis which includes
decays with photons in the final state and report the first observations of
the decays $\bsdsstpi$, $\bsdsrho$ and $\bsdsstrho$.
The leading process of these three modes is a $b\to c\bar ud$
tree-level transition of order $\lambda^2$ with a spectator $s$ quark.
The $\ds$ mesons are reconstructed via three modes :
$\ds\to\phi(\to K^+K^-)\pi^-$, $\ds\to K^{\ast0}(\to K^+\pi^-)K^-$ and
$\ds\to\KS(\to\pi^+\pi^-)K^-$.
Based on the ratio of the second and the zeroth Fox-Wolfram moments
\cite{PRL_41_1581}, $R_2$,
the $e^+e^-\to u\bar u,d\bar d,s\bar s,c\bar c$ continuum events are efficiently
rejected by taking advantage of
the difference between their event geometry (jet like, high $R_2$) and
the signal event shape (spherical, low $R_2$). The $\bsdsstpi$ ($\bsdsrho$ and $\bsdsstrho$) candidates with $R_2$ smaller than 0.5 (0.35) are kept for further analysis.
A best candidate selection, based on the intermediate particle reconstructed masses, is then implemented to keep only one $\bs$ candidate per event.
The $\mbc$ and $\deltae$ distribution of the selected candidates for the three
modes are shown in Fig.~\ref{fig:dominant}, where the various components of the fit are described.

The numerical results are summarized in Table~\ref{tab:dom}.
The systematic error coming from $f_s$ is separated from the other source of
uncertainties.
The $\bsdsstrho$ mode is the decay of the spin-less $\bs$ to two spin-1 particles.
The proportion of longitudinal polarization, $f_L$,  is not known, and the efficiencies depend significantly on this parameter.
A 13\% relative error, denoted ``pol.'', is added to the branching fraction uncertainty to take this into account.

\begin{table}[!h]
\centering
\begin{footnotesize}
\begin{tabular}{l|ccrc}
Mode&$N_{\rm sig}^{\bsst\barbsst}$&$\sum\varepsilon\BR$ ($10^{-3}$)&$S$&Branching fraction\\

\hline
$\bsdsstpi$&$53.4^{+10.3}_{-9.4}$&$9.1\pm0.6$&$8.4\sigma$&$\bfbstodsstpi$\\
$\bsdsrho$&$92.2^{+14.2}_{-13.2}$&$4.4\pm0.3$&$10.6\sigma$&$\bfbstodsrho$\\
$\bsdsstrho$&$86.6^{+15.1}_{-14.0}$&$2.7\pm0.2$&$8.7\sigma$&$\bfbstodsstrho$
\end{tabular}
\end{footnotesize}
\caption{\label{tab:dom}Signal yields for the $\bsst\barbsst$ production process with statistical error, $N_{\rm sig}$,
  total efficiencies, $\sum\varepsilon\BR$,
  statistical significances, $S$,
  and measured branching fractions for the $\bsdsstpi$, $\bsdsrho$ and $\bsdsstrho$ modes.}
\end{table}

These first observations confirm the large predominance of the $\bsst\barbsst$
production mode as no significant excess is seen in the other two signal
regions.
Within uncertainties, no deviation from the $SU(3)$-related $B^0$ decays is
seen.
\section{First observation of the $\mathbf{CP}$-eigenstate decay
$\mathbf{\bsjpsieta}$}

\begin{figure}[!h]
\centering
\begin{minipage}{0.45\linewidth}
\centering
\includegraphics[height=3.8cm]{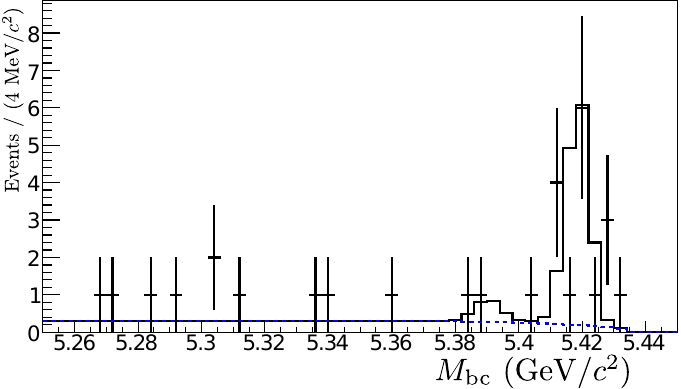}
\end{minipage}~~~
\begin{minipage}{0.45\linewidth}
\centering
\includegraphics[height=3.8cm]{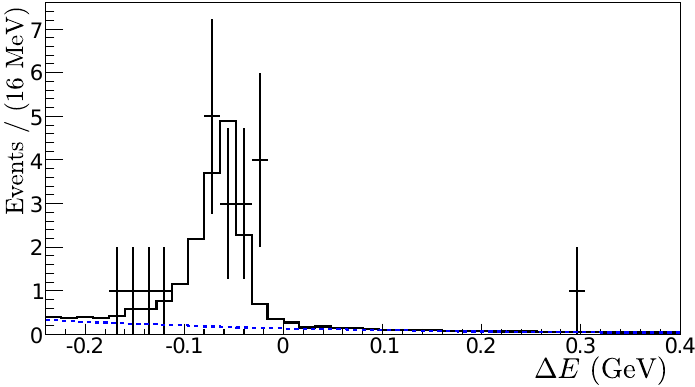}
\end{minipage}
\caption{\label{fig:jpsieta}Projection of the $\bsjpsieta$ candidates
  (points with error bars) and the fitted PDF (solid line)
  in the $\bsst\barbsst$ signal region.
  The sub-modes $\eta\to\gamma\gamma$ and $\eta\to\pi^+\pi^-\pi^0$, which are fitted separately, are summed in these plots.
  The blue-dotted line represents the continuum component of the PDF.
  The small peak in the $\mbc$ plot is the $\bsst\bar\bs$ contribution, as the $\bsst\barbsst$ signal range for $\deltae$ overlaps the one for $\bsst\bar\bs$ signal.
}
\end{figure}

$\bs$ decays to $CP$ eigenstates are important for $CP$-violation parameter measurements
\cite{PRD_63_114015} and preliminary results about the first observation of
$\bsjpsieta$ \cite{hepex_0905_2959v1} are reported.
The $\jpsi$ candidates are formed with oppositely-charged electron or muon pairs,
while $\eta$ candidates are reconstructed via the $\eta\to\gamma\gamma$ and $\eta\to\pi^+\pi^-\pi^0$ modes.
A mass (mass and vertex) constrained fit is then applied to the $\eta$ ($\jpsi$) candidates.
If more than one candidate per event satisfies all the selection criteria, the one with the smallest fit residual is selected.
The main background is the continuum, which is reduced by requiring $R_2<0.4$. 
The combined $\mbc$ and $\deltae$ distributions are presented in Fig.~\ref{fig:jpsieta}.
The $\bs\to\jpsi\,\eta,\,\eta\to\gamma\gamma$ and $\bs\to\jpsi\,\eta,\,\eta\to\pi^+\pi^-\pi^0$ candidates are fitted separately.
Table~\ref{tab:jspieta} presents the numerical results.

\begin{table}[!h]
\centering
\begin{tabular}{l|cc}
& $\bs\to\jpsi\,\eta,\,\eta\to\gamma\gamma$ & $\bs\to\jpsi\,\eta,\,\eta\to\pi^+\pi^-\pi^0$ \\
\hline
$\bsst\barbsst$ Signal Yield&$12.1\pm3.8$&$5.6\pm2.5$\\
Stat. Significance&5.9$\sigma$&4.0$\sigma$\\
\hline
$\BR(\bsjpsieta)$ (combined)&\multicolumn{2}{c}{$\bfbstojpsieta$}
\end{tabular}
\caption{\label{tab:jspieta}$\bsjpsieta$ results: yields, significances and branching fraction.}
\end{table}

\section{Observation of $\mathbf{\bskk}$ and searches for
  $\mathbf{\bspipi}$, $\mathbf{\bskpi}$ and $\mathbf{\bs\to\KS\KS}$}

Finally, we present our results for the $\bskk$, $\bskpi$, $\bspipi$ and $\bs\to\KS\KS$ charmless decays.
The $\bskk$ mode is particularly interesting because it can be used for the determination of
the CKM angle $\gamma$ \cite{PLB_459_306} and may be sensitive to
New Physics \cite{PRD_70_031502}.
The charged pion and kaon candidates are selected using charged tracks and identified with energy deposition, momentum and time-of-flight measurements.
The $\KS$ candidates are reconstructed via the $\KS\to\pi^+\pi^-$ decay, by selecting two oppositely-charged tracks matching various geometrical requirements \cite{phd_ffang}.
A likelihood based on a  Fisher discriminant using 16 modified Fox-Wolfram moments \cite{PRL_91_261801} is implemented to reduce the continuum, which is the main source of background.

\begin{figure}[!h]
\centering
\begin{minipage}{0.3\linewidth}
\centering
\includegraphics[height=4cm]{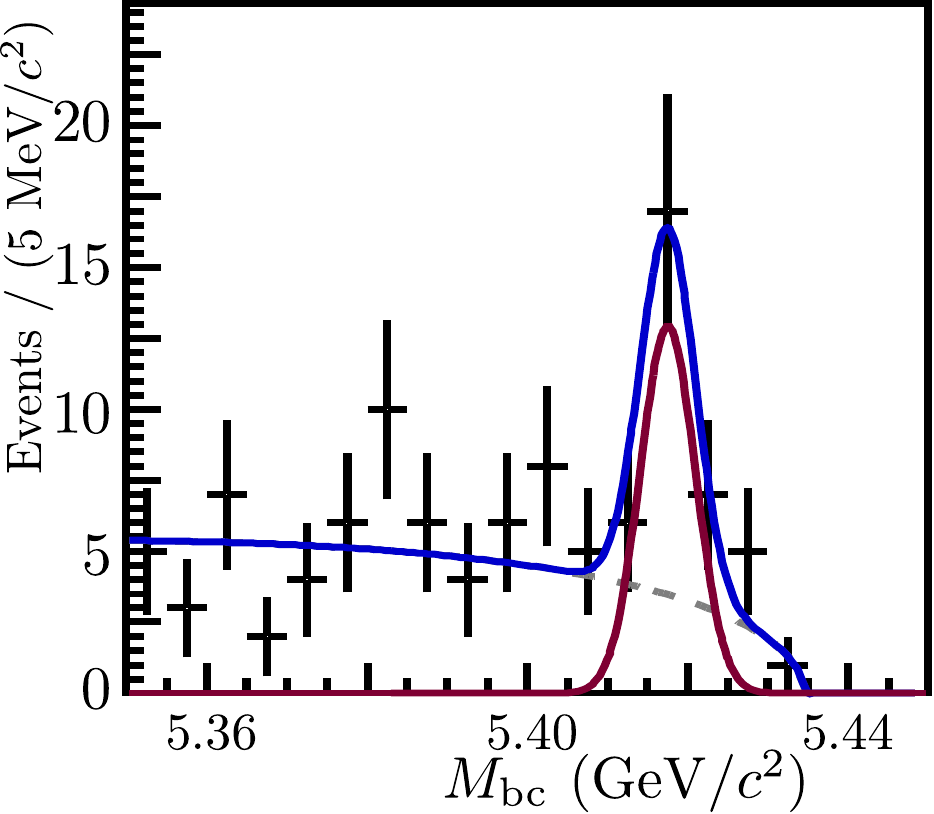}
\end{minipage}~
\begin{minipage}{0.3\linewidth}
\centering
\includegraphics[height=4cm]{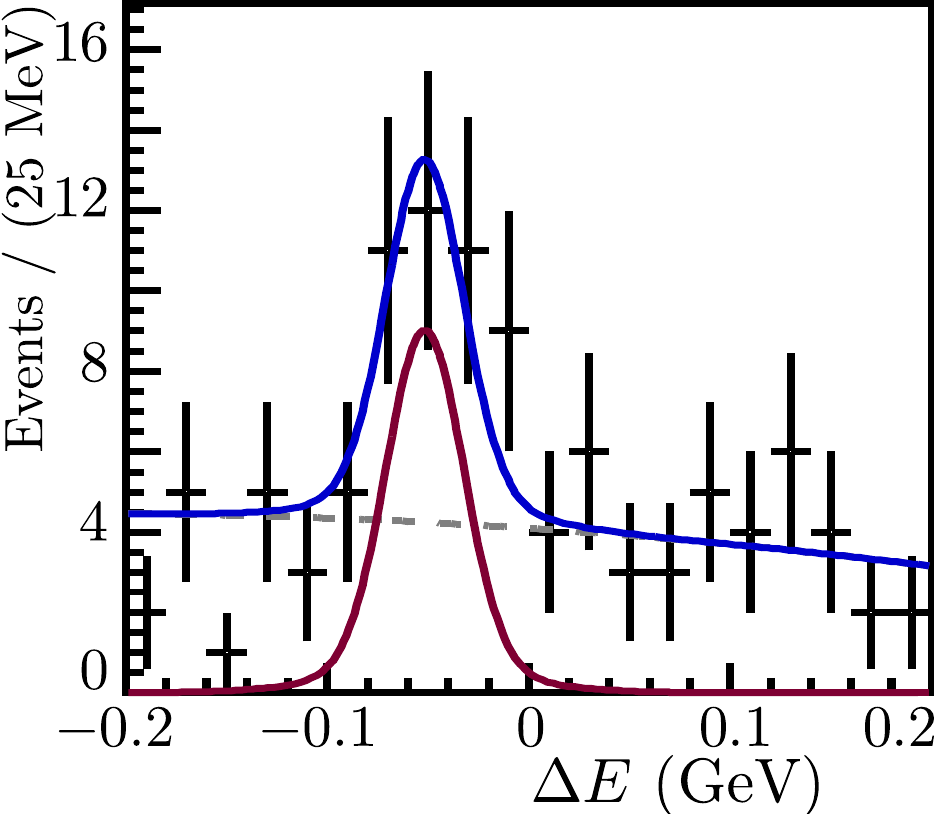}
\end{minipage}~~~~
\begin{minipage}{0.3\linewidth}
\caption{\label{fig:kk}Projections of the $\bskk$ candidates
  (points with error bars) and the fitted PDF (solid blue line)
  in the $\bsst\barbsst$ signal region.
  The solid-red (dotted-\-grey) line represents the signal (continuum)
  component of the PDF.}
\end{minipage}
\end{figure}

We do observe a 5.8$\sigma$ excess of $24\pm6$ events in the $\bsst\barbsst$ region for the $\bskk$ mode (Fig.~\ref{fig:kk}).
The branching fraction $\BR(\bskk)=\bfbstokk$ is derived.
On the other hand, no significant signal is seen for the other modes.
Including the systematics uncertainties, we set the following upper limits at 90\% confidence level:
$\BR(\bspipi)<\bfbstopipi$, $\BR(\bskpi)<\bfbstokpi$ and, assuming $\BR(\bs\to\KS\KS)=\BR(\bs\to\KL\KL)$, $\BR(\bskzkz)<\bfbstokzkz$.
The later is the first limit set for the $\bskzkz$ mode.
All the other values are compatible with the CDF results \cite{PRL_97_211802,PRL_103_031801}.

\section*{Conclusion}

All these studies demonstrate the great potential of
the Belle data set recorded at $\FiveS$ energy.
The sensitivity obtained for several $\bs$ modes allows many interesting studies.
The branching fraction determinations of several important modes is ongoing and eight new measurements have been reported here.
So far, the full Belle sample has reached 100 $\invfb$,
and the KEKB collider may continue delivering collisions
at the $\FiveS$ energy during the fall 2009.
Of course, many more interesting results are
expected with the full Belle $\FiveS$ data set.

\bibliography{bib}{}
\bibliographystyle{JHEP}

\end{document}